\documentclass[lettersize,journal]{IEEEtran}
\IEEEoverridecommandlockouts
\usepackage[hyphens]{url}
\usepackage{breakurl}
\usepackage{cite}
\usepackage{citesort}
\usepackage{amsmath,amsfonts,amsthm,amssymb}
\usepackage{bbm}
\usepackage{algpseudocode}
\usepackage{algorithm}
\allowdisplaybreaks
\usepackage{graphicx}
\usepackage{subfigure}
\graphicspath{{figure/}}
\usepackage{textcomp}
\usepackage{booktabs}
\usepackage{setspace}
\usepackage{tabularx}

\newcolumntype{L}{>{\hspace*{-\tabcolsep}}l}
\newcolumntype{R}{c<{\hspace*{-\tabcolsep}}}
\usepackage[table]{xcolor}
\usepackage{xcolor}
\usepackage{extarrows}
\usepackage{times}
\usepackage[justification = centering]{caption}
\definecolor{lightblue}{rgb}{0.93,0.95,1.0}

\def\BibTeX{{\rm B\kern-.05em{\sc i\kern-.025em b}\kern-.08em
		T\kern-.1667em\lower.7ex\hbox{E}\kern-.125emX}}

\newcommand{\cG}{\mathcal{G}}

\newcommand{\cK}{\mathcal{K}}

\newcommand{\CN}{\mathcal{CN}}

\newcommand{\bh}{\mathbf{h}}

\newcommand{\br}{\mathbf{r}}
\newcommand{\bs}{\mathbf{s}}

\newcommand{\bw}{\mathbf{w}}
\newcommand{\bx}{\mathbf{x}}

\newcommand{\bH}{\mathbf{H}}
\newcommand{\bI}{\mathbf{I}}

\newcommand{\bW}{\mathbf{W}}
\newcommand{\bX}{\mathbf{X}}

\newcommand{\bbR}{\mathbb{R}}

\newcommand{\bbZ}{\mathbb{Z}}

\newcommand{\bzero}{\mathbf{0}}
\newcommand{\bone}{\mathbf{1}}

\newcommand{\figref}[1]{Fig.~\ref{#1}}
\newcommand{\secref}[1]{Section~\ref{#1}}
\newcommand{\subsecref}[1]{Subsection~\ref{#1}}

\newcommand{\lemmref}[1]{\textit{Lemma}~\ref{#1}}

\newcommand{\trace}[1]{\mathrm{tr}\left(#1\right)}
\newcommand{\expect}[1]{\mathbb{E}{\left\{#1\right\}}}

\newtheorem{lemma}{Lemma}
\newtheorem{remark}{Remark}

\begin{document} 
\captionsetup{justification=raggedright,singlelinecheck=false}
\title{Pinching-Antenna-based Communications: Spectral Efficiency Analysis and Deployment Strategies} 
\author{
	Mengyu~Qian, \IEEEmembership{Graduate Student Member,~IEEE,} 
	Xidong~Mu,~\IEEEmembership{Member,~IEEE,} 
	Li~You,~\IEEEmembership{Senior Member,~IEEE,}
    and Michail~Matthaiou,~\IEEEmembership{Fellow,~IEEE}
	\thanks{Mengyu Qian and Li You are with the National Mobile Communications Research Laboratory, Southeast University, Nanjing 210096, China, and also with the Purple Mountain Laboratories, Nanjing 211100, China (e-mail: \{qianmy, lyou\}@seu.edu.cn).}
	\thanks{Xidong Mu and Michail Matthaiou are with the Centre for Wireless Innovation (CWI), Queen’s University Belfast, BT3 9DT Belfast, U.K. (e-mail:  \{x.mu, m.matthaiou\}@qub.ac.uk).}
\thanks{The work of M. Matthaiou was supported by a research grant from the Department for the Economy Northern Ireland under the US-Ireland R\&D Partnership Programme and by the European Research Council (ERC) under the European Union's Horizon 2020 research and innovation programme (grant agreement No. 101001331).}
} 
\maketitle
\begin{abstract}
A multiple-waveguide pinching-antenna (PA)-based multi-user communication system is investigated. With a given number of PAs, two deployment strategies are considered, namely the \emph{centralized PA deployment}, where all PAs are switched between waveguides to serve users in a time-division manner to avail of beamforming gain, and the \emph{distributed PA deployment}, where a single PA is deployed on each waveguide to simultaneously serve multiple users by leveraging the multiplexing gain. The spectral efficiency (SE) achieved by each deployment strategy is analyzed: \romannumeral1) For the centralized deployment,  the positioning strategy of PAs on each waveguide is determined first with the aim of maximizing the channel gain of the corresponding nearest served user.  Based on this, the corresponding system SE is derived. \romannumeral2) For the distributed deployment, the system SE under the maximum ratio transmission (MRT) is first obtained,  which is challenging to analyze due to the inter-user interference. To obtain an analytically tractable form, the stationary phase method is utilized to approximate the system SE. The approximation result reveals that the average inter-user interference can be negligible with a large waveguide spacing and thus the simple MRT is appealing for PA-based multi-user communications. Furthermore, the system SEs achieved by the two strategies are compared in both the high and low signal-to-noise ratio (SNR) regimes. Our analysis suggests that at high SNRs, the distributed deployment is superior to achieve the maximal system SE, while the centralized deployment is more suitable for the low-SNR regime. Finally, the theoretical analysis is verified through simulations. Numerical results demonstrate that \romannumeral1) with the aid of the stationary phase method, the approximation of the system SE proves to be highly accurate overall;  and \romannumeral2)  compared to the centralized deployment, the distributed deployment generally achieves a higher SE but at the cost of a low energy efficiency.
\end{abstract}


\section{Introduction}\label{sec: introduction}
The continuous advancement of wireless communications has spurred the development of many emerging technologies, including reconfigurable intelligent surfaces (RISs), fluid antennas (FAs), movable antennas (MAs), just to name a few \cite{9770295}. Specifically, RISs dynamically manipulate the wireless propagation environment by controlling the phase shifts of reflected signals \cite{liu2021reconfigurable}, and their operational mechanisms as well as applications in communications have been relatively well studied. In contrast, FAs/MAs are relatively new concepts \cite{wong2020performance,ye2025specific}. Generally speaking, FAs/MAs can achieve signal phase adjustment by reconfiguring the physical positions of antennas within the array \cite{zhu2024movable,9264694}.
These technologies share a common mechanism: they enhance the system performance through modifications of small-scale fading characteristics, thereby optimizing transmission reliability, coverage, spectral efficiency (SE) and other performance metrics in next-generation networks.  However, their ability to reduce large-scale path loss is limited, as FAs/MAs only adjust the antenna positions within a range of several wavelengths, while RIS-based transmissions may suffer even more severe large-scale path losses due to the double-fading effect \cite{xu2025pinching,bozanis2025cram}.
To address this issue, the novel pinching-antenna (PA) technology demonstrates unique capabilities in combating large-scale path loss and effectively adjusting small-scale fading \cite{ding2025flexible,Samy2025pinching}, and, for these reasons,  has attracted substantial research interest over the past years.

PA systems are implemented by strategically deploying small dielectric particles, namely PAs, to selectively excite discrete points along a waveguide \cite{xu2025rate}.  Such design effectively transforms the structure into a leaky-wave antenna, since signals will radiate from different PAs on the waveguide \cite{yang2025pinching}.  Since the cost of PAs is relatively low, it is possible to deploy or remove PAs from the waveguide at will, which empowers PA systems to achieve excellent system adaptability.
Moreover,  the waveguide can be extended to considerable lengths, which allows PAs to be positioned in close proximity to users and to create strong line-of-sight paths \cite{liu2025pinching}. 
Therefore, a slight change in the locations of the PAs on the waveguide can lead to a significant variation in the signal received by the user. This is because not only the signal amplitude is affected by the distance between the PAs and the user, but also the signal phase experiences significant variations \cite{zhou2025channel}. As a result, PA systems provide enhanced flexibility for communication designs.

\subsection{Prior Works} 
Given the above mentioned attractive benefits, there are growing research efforts in studying PA-based wireless communications.
The authors of \cite{ding2025flexible} compared the performance gains of PA systems over conventional antenna systems and also proposed a scheme combining PAs with non-orthogonal multiple access (NOMA) to enable multi-user communications via a single waveguide.  This seminal work has inspired numerous follow-up studies focusing on the optimization of the locations of PAs on the waveguide to improve the communication performance. 
For example, 
the authors of \cite{10909665} proposed an approach that separately optimizes the positions of
the PAs and the resource allocation to enhance the uplink performance of  PA-based multi-user communication systems, by maximizing the minimum achievable data rate between devices. 
The authors of \cite{wang2025antenna} investigated NOMA-assisted downlink PA systems and maximized the sum rate of the system by selecting the PAs to be activated on the waveguide. 
The authors of \cite{10981775} analyzed the array gain achieved by PA systems and proposed a PA position refinement method to achieve the upper bound of the array gain. Their work revealed that by optimizing the number of antennas and inter-antenna spacing, PA systems can achieve larger array gain than conventional-antenna systems. In addition, the authors of \cite{tyrovolas2025performance} characterized the optimal placement of PAs to maximize performance under waveguide losses. Further, the authors of \cite{xu2025joint} maximized the system rate by jointly optimizing the transmit and pinching beamforming.
Finally, the authors of \cite{zhao2025waveguide} proposed the new concept of waveguide division multiple access, where by adjusting the activation locations of PAs on each waveguide, the pinching beamforming can be exploited for enhancing the intended user's signal and mitigating the inter-user interference.

\subsection{Motivations and Contributions}
While existing works have undoubtedly revealed the advantages of PA technologies over conventional antenna technologies,  the selection of the of PAs' locations on the waveguide plays a critical role in their practical deployment. For example, for the single-waveguide PA-based systems in \cite{ding2025flexible,performancePASS2025,10981775,cheng2025performancePASS}, it has been demonstrated that a high beamforming gain can be exploited to serve a single user via carefully selecting the PAs' positions. Nevertheless, since a single waveguide can only be fed with the same signal source, multiple-waveguide PA-based communication architectures were further investigated in \cite{xu2025joint,wang2025antenna,ding2025blockage}. With the PAs activated on the designed positions of different waveguides, the multiplexing gain can be exploited to serve multiple users simultaneously.  
It is worth noting that these studies have only focused on selecting the positions of PAs in either single-waveguide or multiple-waveguide PA-based communication systems. However, the deployment strategies of PAs from a system-level perspective remain unexplored. In a general multiple-waveguide PA-based multi-user communication system with a given number of PAs, two typical deployment strategies can be adopted to support multi-user services: \emph{centralized PA deployment}, where all PAs are switched between waveguides to serve users in a time-division manner with enhanced beamforming gain, 
and \emph{distributed PA deployment}, where a single PA is deployed on each waveguide to simultaneously serve multiple users by leveraging multiplexing gain.
It can be observed that with the centralized deployment, each user benefits from a dedicated high beamforming gain but without having the multiplexing gain. In contrast, the distributed deployment offers multiplexing gain by serving multiple users simultaneously. However, each user experiences a reduced beamforming gain and more sophisticated baseband signal processing is generally required to mitigate inter-user interference. To the best of our knowledge, it remains unclear which of these two PA deployment strategies is superior for multi-user PA-based communication systems. This provides the main motivation of this work.

To study the optimal deployment strategies of PAs,  we consider a general PA-based multi-user communication system with a fixed number of PAs. Under the centralized deployment and the distributed deployment strategies, we analyze and evaluate the achieved system SE. 
Our contributions are summarized as follows:
\begin{itemize}
  \item We investigate a general multiple-waveguide PA-based multi-user communication systems, where the base station (BS) is equipped with a PA array and multiple single-antenna users are served through a fixed number of PAs. 
      Then, the system SEs under the centralized PA deployment and the distributed PA deployment are analyzed, respectively.  
  \item We first study the system SE under the centralized PA deployment, where the locations of PAs on the waveguide are determined so that the channel gain of its served user is maximized. Based on this, the system SE can be directly obtained. We then study the system SE under the distributed PA deployment strategy, for which the PA on each waveguide is deployed at the position that is nearest to its served users. Then, the system SE is obtained by utilizing a maximum ratio transmission (MRT) beamformer, where the inter-user interference term complicates the analysis. We utilize the stationary phase method to approximate the system SE. This methodology reveals that a larger waveguide spacing will render the inter-user interference negligible, and therefore the simple MRT scheme is sufficient.
  \item We compare the system SEs  under the centralized PA deployment and the distributed PA deployment  in the high and low  signal-to-noise-ratio (SNR) regimes, respectively.  Our analysis reveals that the distributed PA deployment with high multiplexing gain is preferable in the high-SNR regime to achieve the maximum system SE, while the centralized deployment with high beamforming gain should be adopted in the low-SNR regime.
  \item Finally, our analytical results are verified through simulations. We confirm that the approximation of the system SE under MRT is tight.
  With a practical waveguide spacing, the MRT beamformer proves to be sufficient to maximize the system SE. However, while increasing the waveguide spacing helps reduce the average inter-user interference, it might degrade the SE due to the increased signal propagation path loss.
      Furthermore, the distributed deployment is generally the best choice in terms of achieving a higher system SE compared to the centralized deployment, but this gain comes at the cost of a lower energy efficiency (EE).
\end{itemize}

\subsection{Organization and Notations}
The rest of the paper is organized as follows:
\secref{sec:system model} presents the system model for the PA multi-user communication systems and the system SE achieved by the centralized deployment and the distributed deployment, respectively.
\secref{sec: analysis of system SE} studies analytically the system SE under the centralized PA deployment first. Then, the analysis of the system SE under the distributed PA deployment is provided.
\secref{sec: comparison_SE_special_cases} compares the SE of the two deployment strategies in both low- and high-SNR regimes.
\secref{sec: numerical results}  provides simulation results to validate the theoretical analysis. 
Finally, \secref{sec: conclusions} concludes the paper.

\emph{Notations:} Scalars, vectors, and matrices are denoted by
lower-case, bold-face lower-case, and bold-face upper-case letters, respectively;
 $\bX^T$ and $\bX^H$ are the transpose and conjugate transpose of $\bX$, respectively;  $\bx \succeq \bzero$ indicates that each element of $\bx$ is non-negative; $||\bx||$ is the Euclidean norm and $|\bx|$ represents the modulus. Calligraphic letters, e.g., $\cG$, denote sets, with $|\cG|$ being the set cardinality;
 $\expect{\cdot}$ and $\trace{\cdot}$ are the expectation operator and trace operator, respectively; $\mathcal{CN}(a,b)$ represents a Gaussian distribution, with $a$ being the mean and $b$ being the variance. Finally, $\bone_N$ is the $N \times 1$ vector with all the elements being $1$.

\section{System Model}\label{sec:system model}
\begin{figure}[t]
	\centering
    \captionsetup{font={small}}
    \includegraphics[width=0.47\textwidth]{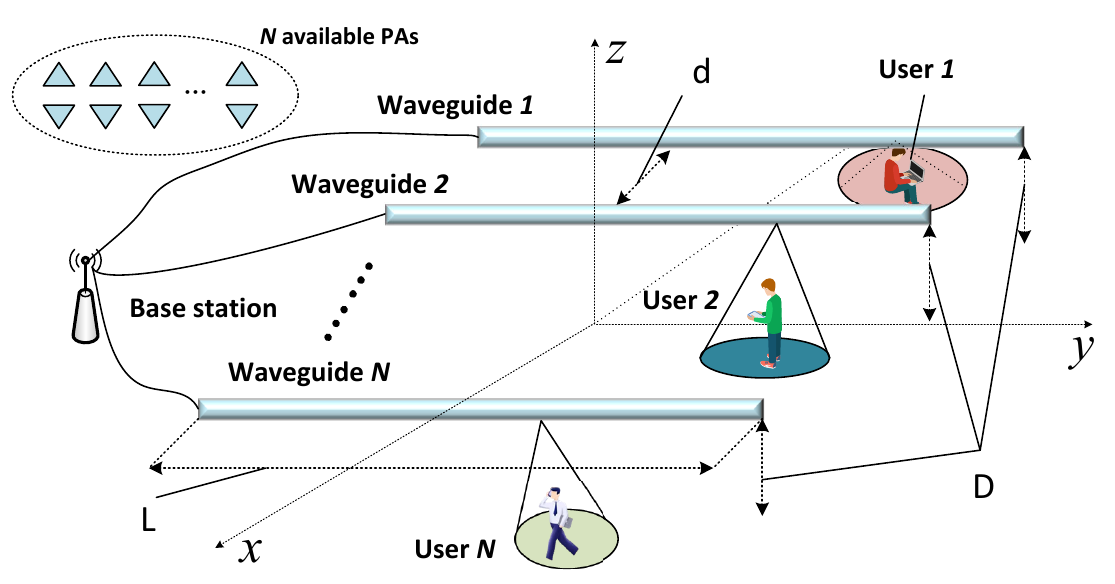}
	\caption{The general layout of the considered PA-based multi-user communication system.}
    \label{PASS}
\end{figure}

In this section, we present the system model of the studied PA-based multi-user communication system and the two PA deployment strategies.
As shown in \figref{PASS}, the BS is connected with  $N$ waveguides to serve $N$ users with $N$ PAs available. A three-dimensional (3D) Cartesian coordinate system is considered, where all the waveguides are aligned parallel to the $y$ axis with the same height $D$. The spacing between adjacent waveguides is denoted  by $d$.
The length of each waveguide is denoted by $L$.
The location of the $q$th PA activated on the $i$th waveguide, i.e., the $(i,q)$th PA, $\forall i,q = \{1,2,\dots, N\}$, is represented by $\bs_{i,q}  = [\bar{x}_i,y_{i,q},D]^T$, where $-L/2 \leq  y_{i,q} \leq L/2$, while $\bs_{i,0} = [\bar{x}_i,-L/2,D]^T$ is the location of the feed point of the $i$th waveguide. 
Meanwhile, let $\cK \triangleq \{1,2,\dots,N\}$ denote the user set. The location of the $k$th single-antenna user is denoted by $\br_k = [x_k,y_k,0]^T,\forall k \in \cK$.  Note that the width of the waveguide is negligible compared to the waveguide spacing $d$, while the $x$-coordiante of the first waveguide is assumed to be $\bar{x}_1 = 0$ for simplicity. Then, the $x$-coordinate of the $i$th waveguide is
 \begin{align}
  \bar{x}_i = (i-1)d,\quad i=\{1,2,\dots,N\}.
 \end{align}
Assume that user $k \in\cK$ is only closest to one specific waveguide, and for notational simplicity, let the waveguide closest to the $k$th user share the same index  $k$ with the served user. Moreover, we assume that $x_k$ is uniformly distributed among the range of $[kd-\frac{3d}{2},kd-\frac{d}{2}],\forall k \in\cK$, with the possibility being $1/d$.
 
From a system design perspective, we propose two PA deployment strategies to provide multi-user service, namely the \emph{centralized PA deployment} and the \emph{distributed PA deployment}. Specifically, for the centralized PA deployment, all the $N$ PAs are switched between waveguides to serve $N$ users in a time-division manner, which harvests the maximal beamforming gain for each user. For the distributed PA deployment, 
the $N$ PAs are deployed on $N$ waveguides to serve $N$ users simultaneously, which harvests the multiplexing gain.
In the following, we will establish the system models along with the system SEs for the two PA deployment strategies.

\subsection{Centralized PA Deployment}
With the centralized deployment, the BS communicates with the $N$ users simultaneously in $N$ orthogonal time slots of equal size. In each time slot, 
all the $N$ PAs are deployed on one specific waveguide to serve its nearest user. 

In this case, the received signal of the served user $k \in \cK$ in the served time slot is given by \cite{ding2025flexible}
\begin{align}
  y_k = \sqrt{\frac{P_t}{N}}\sum_{q=1}^{N} h(\br_k,\bs_{k,q}) x_k + n_k,
\end{align}
where $x_k,\forall k \in \cK$, is the signal transmitted to user $k$, which satisfies $\expect{x_k} = 0$ and $\expect{|x_k|^2} = 1$; $P_t$ is the total transmit power for serving the $N$ users; $n_k$ is the noise at the receiver with the distribution being $n_k \sim \CN(0,\sigma^2)$;
$h(\br_k,\bs_{k,q})$ represents the channel response between the $(k,q)$th PA and user $k$, which is given by
\begin{align}
  h(\br_k,\bs_{k,q}) = \frac{\sqrt{\eta} e^{-\jmath (\frac{2\pi}{\lambda}\left|\br_k-\bs_{k,q}\right|+\frac{2\pi}{\lambda_g}\left|\bs_{k,q}-\bs_{k,0}\right|)}}{\left|\br_k-\bs_{k,q}\right|},
\end{align}
where $\eta = (\lambda/4\pi)^2$ represents the channel gain at the reference distance of 1 m  with $\lambda$ being the wavelength; $\lambda_g \triangleq \lambda/n_{\rm eff}$, with $n_{\rm eff}$ being the effective refractive index of a dielectric waveguide \cite{zhao2025waveguide}; $|\br_k-\bs_{k,q}|,\forall k \in \cK, q=\{1,2,\dots,N\}$ denotes the distance between user $k$ and the $q$th PA on the $k$th waveguide, while $|\bs_{k,q}-\bs_{k,0}|,\forall k \in \cK, q=\{1,2,\dots,N\}$ denotes the propagation distance from the feed point to the $q$th PA on the $k$th waveguide.

As there is only one user served in each time slot, the SNR of user $k$ is given by
\begin{align}
  {\rm SNR}_k = \frac{\frac{P_t}{N}\left|\sum_{q=1}^{N} h(\br_k,\bs_{k,q})\right|^2}{\sigma^2}.
\end{align}

Due to the time-divison operation mode, the achievable SE of user $k,\forall k \in \cK$, is given by
\begin{align}
  R_k =& \frac{1}{N}\log_2\left(1+ {\rm SNR}_k\right)\notag\\
  =& \frac{1}{N} \log_2\left(1+ \frac{\frac{P_t}{N}\left|\sum_{q=1}^{N} h(\br_k,\bs_{k,q})\right|^2}{\sigma^2}\right).
\end{align}
 Thus, the system SE under the centralized deployment is
\begin{align}\label{SE_centralized_original}
  C_{\rm c} &= \sum_{k=1}^{N}R_k \notag \\
  &= \frac{1}{N} \sum_{k=1}^{N} \log_2\left(1+ \frac{\frac{P_t}{N}\left|\sum_{q=1}^{N} h(\br_k,\bs_{k,q})\right|^2}{\sigma^2}\right).
\end{align}

\subsection{Distributed PA Deployment}
With the distributed PA deployment, a single PA is activated on a dedicated waveguide. Therefore, all $N$ waveguides are activated to radiate signals to $N$ users simultaneously, where different waveguides can be fed with different signals.
Therefore, the received signal of user $k \in \cK$ is \cite{ding2025flexible}
\begin{align}
  y_k = &\sum_{i=1}^{N}\sqrt{p_k} h(\br_k,\bs_{i})  w_{k,i} x_k \notag \\
  &+ \sum_{k' \neq k}^{N}\sum_{i=1}^{N}\sqrt{p_{k'}} h(\br_k,\bs_{i})  w_{{k'},i} x_{k'}+ n_k,
\end{align}
where $x_k, \forall k \in \cK$, is the signal transmitted to user $k$. Denote the signal vector as $\bx = [x_1,x_2,\dots,x_N]^T$, which satisfies $\expect{\bx} = \bzero$  and $\expect{\bx \bx^H} =  \bI_N$; $\bw_k = [w_{k,1},w_{k,2},\dots,w_{k,N}]^T,\forall k \in \cK$, includes the beamforming coefficients assigned to the signal transmitted to user $k$, with $w_{k,n},\forall k \in \cK,n=\{1,2,\dots,N\}$, being the beamforming coefficient assigned to the signal of user $k$ on the $n$th waveguide. Further,
$h(\br_k,\bs_{i})$ is the channel coefficient between user $k$ and the PA on the $i$th waveguide, 
which is given by
\begin{align}
  h(\br_k,\bs_{i}) = \frac{\sqrt{\eta} e^{-\jmath (\frac{2\pi}{\lambda}\left|\br_k-\bs_{i}\right|+\frac{2\pi}{\lambda_g}\left|\bs_{i}-\bs_{i,0}\right|)}}{\left|\br_k-\bs_{i}\right|}.
\end{align}
Here, we have simplified the representation of $\bs_{i,q}$ by $\bs_{i}$, since only one PA is on the $i$th waveguide. Finally,
$p_k,\forall k \in \cK,$ represents the power transmitted to user $k$, with the total transmit power being $\sum_{k=1}^{N} p_k = P_t$. 

Then, the signal-to-interference-plus-noise-ratio (SINR) of user $k$ is given by
\begin{align}
  {\rm SINR}_k &= \frac{p_k\left|\sum_{i=1}^{N} h(\br_k,\bs_i)  w_{k,i} \right|^2}{\sum_{k' \neq k}^{N} p_{k'} \left|\sum_{i=1}^{N}h(\br_k,\bs_i)  w_{k',i}\right|^2+\sigma^2}\notag \\
  &= \frac{p_k \left|\bw_k^T\bh(\br_k,\bs)\right|^2}{\sum_{k' \neq k}^{N} p_{k'}\left|\bw_{k'}^T\bh(\br_k,\bs)\right|^2 +\sigma^2},
\end{align}
where $\bh(\br_k,\bs) \triangleq [h(\br_k,\bs_{1}),h(\br_k,\bs_{2}),\dots,h(\br_k,\bs_{N})]^T$ represents the channel vector between the  $N$ PAs on the $N$ waveguides and user $k$, where $\bs \triangleq [\bs_1,\bs_2,\dots,\bs_N] \in \bbR^{3 \times N}$. 

Correspondingly, the system SE under the distributed PA deployment is given by
\begin{align}\label{SE_distributed_original}
  C_{\rm d} =& \sum_{k=1}^{N} \log_2\left(1+ {\rm SINR}_k\right) \notag \\
  =& \sum_{k=1}^{N} \log_2\left(1+ \frac{p_k \left|\bw_k^T\bh(\br_k,\bs)\right|^2}{\sum_{k' \neq k}^{N} p_{k'}\left|\bw_{k'}^T\bh(\br_k,\bs)\right|^2 +\sigma^2}\right).
\end{align}

\subsection{Discussion}
It can be observed that the system SE in \eqref{SE_centralized_original} under the centralized PA deployment benefits from a high beamforming gain and interference-free transmission. In contrast, the system SE in \eqref{SE_distributed_original} under the distributed PA deployment achieves the highest multiplexing gain, albeit at the cost of inter-user interference. While both the beamforming and multiplexing gains have the potential to improve the system SE, it is unknown which of the two strategies is better.
Furthermore, the system SE under the distributed deployment is determined by both the baseband beamformers and the locations of PAs, while in the centralized deployment, it depends solely on the locations of PAs.

\section{SE Analysis for Centralized and Distributed PA Deployments}\label{sec: analysis of system SE}
In this section,  we will analyze the system SE under the centralized deployment and the distributed deployment, respectively. For each case, we will determine the locations of the PAs on the waveguides first, and then, derive the analytically tractable form of the system SE.

\subsection{SE Analysis with the Centralized PA Deployment}\label{SE_Analysis_Centralized}
In the centralized PA deployment strategy, all $N$ PAs are deployed on one waveguide during a given time slot and then switched to another waveguide in the following time slot, successively serving $N$ users.  As the locations of PAs on the waveguide affect not only the amplitude of the radiated signal but also the phase of the signal, it is important to determine the locations of PAs.   Thus, we determine the deployment of PAs on each serving waveguide first. Then, we will derive the system SE under this location strategy.

\subsubsection{Locations of PAs}
To maximize the received SNR at the users, we determine the locations of PAs on the serving waveguide so that the signals received by the user are combined in-phase \cite{performancePASS2025}. 
For notational simplicity, $\forall k \in\cK, q = \{1,2,\dots,N\}$, let
\begin{subequations}
\begin{align}
  L_{k,q} & \triangleq |\bs_{k,q}-\bs_{k,0}| = y_{k,q}+\frac{L}{2},  \\
  d_{k,k,q} & \triangleq |\br_k-\bs_{k,q}| = \sqrt{(x_k-\bar{x}_k)^2+(y_k-y_{k,q})^2+D^2}.
\end{align}
\end{subequations}

In order to make sure that the signals received by the user are combined in-phase,  the following equation should be satisfied:
\begin{align}
  d_{k,k,q+1}+n_{\rm eff}L_{k,q+1}- (d_{k,k,q}+n_{\rm eff}L_{k,q}) = z\lambda, z \in \bbZ,
\end{align}
where we have assumed that $y_k-y_{k,q} \ll \sqrt{D^2+(x_k-\bar{x}_k)^2}$. As will be shown later, when serving user $k \in \cK$, the $y$-coordinates of the PAs on the $k$th waveguide are all located near the $y$-coordinate of user $k$,  i.e., $y_k-y_{k,q} \approx 0$. Hence,
the spacing between the adjacent PAs on the $k$th waveguide can be approximated by $\lambda/n_{\rm eff} \triangleq \lambda_e$. 

With the above configuration, we have
\begin{align}\label{h_mi}
  \sum_{q=1}^{N} h(\br_k,\bs_{k,q}) &=\sum_{q=1}^{N} \frac{\sqrt{\eta}e^{-\jmath\frac{2\pi}{\lambda}(d_{k,k,q}+n_{\rm eff}L_{k,q})}}{d_{k,k,q}}\notag \\
  &= e^{-\jmath\phi_{k,k}} \sum_{q=1}^{N} \frac{\sqrt{\eta}}{d_{k,k,q}},
\end{align}
where $e^{-\jmath\phi_{k,k}}$ is the common phase factor.

Denote the PA that determines $e^{-\jmath\phi_{k,k}}$ as the reference PA, with the location of it being the reference point on the waveguide, which will also affect the large-scale path loss $1/d_{k,k,q},\forall k \in \cK,q=\{1,2,\dots,N\}$. To maximize the strength of \eqref{h_mi}, the reference point should be chosen as the location on the waveguide closest to its served user to minimize the large-scale path loss.

Specifically, denote the reference point on the $k$th waveguide as $O_{k} = [\bar{x}_k,y_k,D]^T, k\in \cK$. 
The reference PA is placed at the reference point. 
Then, arrange PAs numbered $1,2,\dots,(N-1)/2$ along the positive direction of waveguide propagation, while arrange PAs numbered $-1,-2,\dots,-(N-1)/2$ along the opposite direction of waveguide propagation. Thus, $y_{k,q} = y_k + q\lambda_e, q = \{-N/2,\dots,0,\dots,N/2\},\forall k \in \cK$.
Here, we have assumed that $N$ is an odd number. If $N$ is even, we can remove the PA at the reference point, and then still denote the antennas along the positive direction of waveguide propagation as $1,2,\dots,N/2$, while $-1,-2,\dots,-N/2$ along the opposite direction of waveguide propagation.

\subsubsection{System SE with the Centralized PA Deployment} 
With the configuration proposed above, the common phase factor is
\begin{align}
  \phi_{k,k} = \frac{2\pi}{\lambda}\sqrt{(x_k-\bar{x}_k)^2+D^2},\quad \forall k \in \cK.
\end{align}
The large-scale path loss, denoted by $1/d_{k,k,q}$, can be approximated as follows:
\begin{align}
  \frac{1}{d_{k,k,q}} \overset{\text{(a)}}{\approx} \frac{1}{d_{k,k}} \triangleq \frac{1}{\sqrt{(x_k-\bar{x}_k)^2+D^2}},\quad \forall k\in \cK.
\end{align}
The approximation in (a) is because  $|y_{k,q}- y_k| =  q\lambda_e,\forall q \in \{1,2,\dots,N\}$, can be negligible due to  $\lambda_e \ll 1$.

Thus, the system SE under the centralized deployment becomes
\begin{subequations}\label{SE_centralized}
\begin{align}
  C_{\rm c} =& \frac{1}{N} \sum_{k=1}^{N} \log_2\left(1+\frac{P_t N \eta }{d_{k,k}^2\sigma^2 }\right)\\
  \overset{\text{(a)}}{\approx} & \log_2\left(1+\frac{P_t N \eta }{d_{k,k}^2\sigma^2 }\right),
\end{align}
\end{subequations}
where the approximation in (a) is due to the fact that the distribution of users around their nearest waveguide is identical. Thus, the value of $d_{k,k},\forall k \in \cK,$ follows a similar distribution for different users.
Moreover, it can be observed that by increasing the number of PAs $N$ on the waveguide, the effective signal strength, i.e., the beamforming gain, will be enhanced.

\subsection{SE Analysis with the Distributed PA Deployment}\label{SE_Analysis_Distributed}
Different from  the centralized deployment, there is only one PA on each waveguide for the distributed deployment, and the $N$ users are served simultaneously. Thus, there exists inter-user interference and the design of the beamformer is required. In the following, we will first determine the locations of the PAs on each waveguide, and then derive the corresponding system SE with the MRT beamformer.

\subsubsection{Locations of PAs}
For the distributed deployment case, where each waveguide is equipped with only one PA, 
we note that the locations of the PAs have a significant impact on the system performance. To facilitate our performance analysis and provide intuitive design insights, a straightforward and effective approach is to deploy each PA at the location closest to its served user. 
Specifically, the location of the PA on the $i$th waveguide is $\bs_i = [\bar{x}_i,y_i,D]^T$, with $0 \leq |x_i -\bar{x}_i|\leq d,i=\{1,2,\dots,N\}$. To maintain consistency with the descriptions before, we also refer to the location denoted by $\bs_i$ as the reference point on the $i$th waveguide.
It is worth noting that this deployment only considers maximizing the received signal strength, without taking inter-user interference suppression into account, and therefore may not be optimal from an optimization perspective. However, the subsequent analysis will show that inter-user interference can be mitigated under certain conditions. 

\subsubsection{System SE with MRT Beamformer}
It is worth noting that while beamformer design is of significant importance in multi-user systems with inter-user interference--as it can substantially suppress interference and enhance the system performance--the optimization of beamforming strategies constitutes a separate and extensive research topic \cite{9035662,9973235,10458417,10505154}. In this study, our objective is to obtain communication system-level insights rather than to optimize the beamformer itself. To this end, we adopt the simple MRT approach, which enables a clearer inspection of the effects of PA deployment, i.e., the channel response characteristics, on the overall system SE.

When utilizing the MRT beamformer, $\bw_k = \frac{\bh^*(\br_k,\bs)}{||\bh(\br_k,\bs)||_2}$, the system SE becomes
\begin{align}\label{SE_distributed_MRT}
  C_{\rm d} = \sum_{k=1}^{N} \log_2\left(1+ \frac{p_k ||\bh(\br_k,\bs)||^2}{\sum_{k' \neq k}^{N} p_{k'}\frac{\left|\bh^H(\br_{k'},\bs)\bh(\br_k,\bs)\right|^2}{||\bh(\br_{k'},\bs)||^2} +\sigma^2}\right).
\end{align}

With the locations of PAs proposed above, the $i$th  element of the channel response $\bh(\br_k,\bs)$ is given by
\begin{align}
  h(\br_k,\bs_i) = \frac{\sqrt{\eta} e^{-\jmath \frac{2\pi}{\lambda}\left[d_{k,i} + n_{\rm eff}(y_i+\frac{L}{2})\right]}}{d_{k,i}}, \forall i = \{1,2,\dots,N\},
\end{align}
where $d_{k,i} = \sqrt{(x_k-\bar{x}_i)^2+(y_k-y_i)^2+D^2}$.

Thus, we have the following result:
\begin{align}\label{hk_hm_squared}
  \left|\bh^H(\br_{k'},\bs) \bh(\br_k,\bs)\right|^2 = \left| \eta \sum_{i=1}^{N}  \frac{e^{-\jmath\frac{2\pi}{\lambda}(d_{k,i}-d_{k',i})}}{d_{k',i}d_{k,i}}\right|^2.
\end{align}
Similarly, we obtain $||\bh(\br_k,\bs)||^2,\forall k \in \cK,$ as follows:
\begin{align}\label{hm_squared}
  ||\bh(\br_k,\bs)||^2 =  \sum_{i=1}^{N} \frac{\eta}{d_{k,i}^2}.
\end{align}

By substituting \eqref{hk_hm_squared} and \eqref{hm_squared} into \eqref{SE_distributed_MRT}, the system SE under the distributed deployment with MRT beamformer can be expressed as follows:
\begin{align}\label{SE_distributed_untractable}
  & C_{\rm d} = \notag \\
 & \sum_{k=1}^{N} \log_2 \left(1+ \frac{p_k \sum_{i=1}^{N} \frac{1}{d_{k,i}^2}}{\sum_{k' \neq k}^{N} p_{k'} \frac{ \left|\sum_{i=1}^{N}  \frac{e^{-\jmath\frac{2\pi}{\lambda}(d_{k,i}-d_{k',i})}}{d_{k',i}d_{k,i}}\right|^2}{\sum_{i=1}^{N} \frac{1}{d_{k',i}^2}} + \frac{\sigma^2}{\eta}}\right).
\end{align}

It can be observed that the strength of the effective received signal is mainly determined by the large-scale path loss, i.e.,  $d_{k,i},\forall k \in \cK, i=\{1,2,\dots,N\}$, which is only dependent on the distance between user $k$ and the reference points on each waveguide.
In contrast, the inter-user interference is not only affected in magnitude by these distance terms, but its phase components also depend on them, which complicates the expression of the interference and makes our analysis more challenging. However, it also offers potential opportunities for suppressing inter-user interference. 
For example, we can reduce the inter-user interference by introducing a small deviation from the reference point, i.e., set the reference point on the $i$th waveguide as $O_i = [\bar{x}_i,y_i+\delta_i,D]^T$, with $\delta_i$ being the small deviation.  
Since $\delta_i,i=\{1,2,\dots,N\}$ is small, its effect on the magnitude term can be safely ignored. On the contrary, by leveraging the characteristic of rapid phase variations, we can optimize $\{\delta_1,\delta_2,\dots,\delta_N\}$ to minimize $\big|\sum_{i=1}^{N} e^{-\jmath\frac{2\pi}{\lambda}(d_{k,i}-d_{k',i})}/(d_{k,i}d_{k',i})\big|^2$ as much as possible.
The specific design process will be briefly described in the following.

The propagation distance after introducing the small deviations $\delta_i,\forall i=\{1,2,\dots,N\}$, can be approximated by
\begin{subequations}
\begin{align}
  \tilde{d}_{k,i}&= \sqrt{(x_k-\bar{x}_i)^2+(y_k-y_i+\delta_i)^2+D^2} \\
  &\approx d_{k,i}+\frac{y_k-y_i}{d_{k,i}}\delta_i,\label{tilde_d_mi_approximation}
\end{align}
\end{subequations}
where the approximation in \eqref{tilde_d_mi_approximation} is because $\delta_i \ll d_{k,i}$, as $\delta_i$ is extremely small. Then, we
denote the interference between user $k$ and user $k'$ as $I_{k,k'}$, which is given by
\begin{align}
  I_{k,k'} = \left|\sum_{i=1}^{N} \frac{e^{-\jmath \frac{2\pi}{\lambda}(\tilde{d}_{k,i}-\tilde{d}_{k',i})}}{d_{k,i}d_{k',i}}\right|^2.
\end{align}

In the ideal case, we expect that $I_{k,k'} = 0,k,k' \in\cK,\forall k \neq k'$.  However, from the perspective of optimization degrees of freedom, achieving $I_{k,k'} = 0, k,k' \in\cK,\forall k \neq k'$, would require satisfying $N(N-1)/2$ constraint conditions while only $N$ variables, i.e., $\{\delta_1,\delta_2,\dots,\delta_N\}$, are available for optimization. Thus, it is possible to minimize the inter-user interference by introducing small deviations when $N(N-1)/2 \leq N$, i.e., the number of users $N \leq 3$ in this system. 
For example, consider $N=2$. 
In this case, one possible set of $\{\delta_1,\delta_2\}$ can be obtained by satisfying the following equation:
\begin{subequations}\label{example_deviation}
  \begin{align}
 &I_{1,2}  = I_{2,1} = \left|\sum_{i=1}^{2} \frac{e^{-\jmath \frac{2\pi}{\lambda}(\tilde{d}_{k,i}-\tilde{d}_{k',i})}}{d_{k,i}d_{k',i}}\right|^2 = 0\\
 \Rightarrow & \frac{e^{-\jmath \frac{2\pi}{\lambda}(\tilde{d}_{1,1}-\tilde{d}_{2,1})}}{d_{1,1}d_{2,1}} + \frac{e^{-\jmath \frac{2\pi}{\lambda}(\tilde{d}_{1,2}-\tilde{d}_{2,2})}}{d_{1,2}d_{2,2}} = 0\label{exact_equations_interference_reduction}.
  \end{align}
\end{subequations}
The exact solutions to \eqref{example_deviation} require numerical computation or even optimization. However, from the perspective of merely reducing inter-user interference, a simple set of solutions can be obtained by letting the two terms in \eqref{exact_equations_interference_reduction} out-of-phase, which leads to
\begin{subequations}
  \begin{align}
 & \tilde{d}_{1,1}-\tilde{d}_{2,1} = \tilde{d}_{1,2}-\tilde{d}_{2,2}+\frac{\lambda}{2}+z\lambda,\quad z \in \bbZ,\\
  \Rightarrow & d_{1,1}+d_{2,2}-d_{1,2}-d_{2,1} = \notag \\
  &\quad \quad \quad \quad \quad \quad \frac{y_1-y_2}{d_{1,2}}\delta_2+\frac{y_2-y_1}{d_{2,1}}\delta_1+\frac{\lambda}{2} + z\lambda\label{eli_interference}.
  \end{align}
\end{subequations}

It can be found that as long as $y_1 \neq y_{2}$ (or in a general case, $y_k \neq y_{k'},k,k' \in \cK, \forall k \neq k'$), we can choose appropriate $\{\delta_1,\delta_2\}$ (or in general case $\{\delta_1,\dots, \delta_N\}$)  to satisfy \eqref{eli_interference}. However, when users are aligned in a straight line along the vertical propagation direction, i.e., the coordinates along the $y$ axis are all the same in this paper, the inter-user interference becomes inherently difficult to reduce as the impact of the small deviations $\{\delta_1,\dots,\delta_N\}$ disappear. From a practical perspective, this condition is extremely unlikely to occur. However, it has theoretical value as it represents the worst-case scenario.

In addition, as mentioned before, when $N \geq 4$, it also becomes hard to eliminate the inter-user interference, since the following equations should be satisfied with only $N$ variables:
\begin{align}
  \left\{
  \begin{aligned}
   & I_{1,2}\left(\delta_1,\dots,\delta_N\right)  =  I_{2,1}\left(\delta_1,\dots,\delta_N\right) = 0,\\
   & I_{1,3}\left(\delta_1,\dots,\delta_N\right) =  I_{3,1}\left(\delta_1,\dots,\delta_N\right) = 0,\\
   & \dots \\
   & I_{N-1,N}\left(\delta_1,\dots,\delta_N\right)  =  I_{N,N-1}\left(\delta_1,\dots,\delta_N\right) = 0.
  \end{aligned}
  \right.
\end{align}
Given this, it is important to investigate the impact of inter-user interference on the system SE, which will be analyzed in the following.

\subsubsection{Analysis of Average Inter-user Interference}
It is worth noting that, in the worst-case scenario, mitigating inter-user interference becomes challenging. Therefore, it is important to analyze its impact under such conditions. In this case, we have
\begin{align}
  d_{k,i} = \sqrt{(x_k-\bar{x}_i)^2+D^2},\quad \forall k\in\cK,i=\{1,2,\dots,N\}.
\end{align}
Then, the average strength of the interference between user $k'$ and user $k$ is denoted by
\begin{align}
  \bar{I}_{k',k} = \frac{1}{d^4}\left|\sum_{i=1}^{N}f(k',i)f^*(k,i)\right|^2,
\end{align}
where
\begin{subequations}
\begin{align}
  f(k',i) &= \int_{k'd-\frac{3d}{2}}^{k'd-\frac{d}{2}}\frac{e^{-\jmath \frac{2\pi}{\lambda}\sqrt{[x-(i-1)d]^2+D^2}}}{\sqrt{[x-(i-1)d]^2+D^2}}  dx\\
  &\triangleq  \int_{k'd-\frac{3d}{2}}^{k'd-\frac{d}{2}}e^{\jmath \phi_i(x)} g_i(x) dx,
\end{align}
\end{subequations}
and $\phi_i(x)\triangleq -\frac{2\pi}{\lambda}\sqrt{[x-(i-1)d]^2+D^2}\triangleq -\frac{2\pi}{\lambda}f_i(x)$ is the phase function, while  $g_i(x)\triangleq 1/\sqrt{[x-(i-1)d]^2+D^2}$ is the amplitude function.

The following lemma provides the approximated strength of the average interference.
\begin{lemma}\label{lemma_1}
  The strength of the average inter-user interference $\bar{I}_{k',k},\forall k'\neq k, k',k\in \cK,$ can be approximated as follows:
\begin{align}\label{I_mk}
  \bar{I}_{k',k} \approx \frac{\lambda^3}{\pi^2 d^6 D} T^2(k',k),
\end{align}
where
\begin{align}
  T(k',k) &= \frac{\cos\left(\frac{2\pi}{\lambda} \sqrt{(k'-k+\frac{1}{2})^2 d^2+D^2}-\frac{2\pi}{\lambda}D-\frac{\pi}{4} \right)}{k'-k+\frac{1}{2}}\notag \\
  & - \frac{\cos\left(\frac{2\pi}{\lambda}\sqrt{(k'-k-\frac{1}{2})^2 d^2+D^2}-\frac{2\pi}{\lambda}D-\frac{\pi}{4} \right)}{k'-k-\frac{1}{2}}.
\end{align}
\end{lemma}
\begin{proof}
Given the fact that the phase function $\phi_i(x)$ will vary rapidly with $x$, while the amplitude term $g_i(x)$ varies slowly within the small range $x\in[k'd-3d/2,k'd-d/2]$,  we adopt the stationary phase point method to approximate the integral \cite{tong2025near}.
First, to find the stationary phase point, we let
\begin{align}
  \frac{d \phi_i(x)}{d x} = 0 \Rightarrow x = (i-1)d.
\end{align}
When $k'd-3d/2\leq x = (i-1)d \leq k'd-d/2$,  it leads to $i = k'$. Thus, $\forall k' \in \cK$,
\begin{align}\label{f_mm}
  f(k',k') &\approx \frac{e^{ -\jmath\frac{2\pi D}{\lambda}}}{D}  \sqrt{\jmath \lambda D} = \sqrt{\frac{\lambda}{D}}e^{ -\jmath(\frac{2\pi D}{\lambda}-\frac{\pi}{4})}.
\end{align}

When $i \neq k'$, the main contribution to the integral comes from the endpoints $x_{lu} \triangleq k'd-3d/2$ and $x_{up} \triangleq k'd-d/2$ \cite{wong2001asymptotic}. Then,
\begin{align}
  f(k',i) &= \frac{e^{-\jmath  \frac{2\pi}{\lambda} f_i(k'd-\frac{d}{2})}}{-\jmath  \frac{2\pi}{\lambda} f_i^{\prime}(k'd-\frac{d}{2})} \sum_{l=0}^L \frac{g_i^{(l)}(k'd-\frac{d}{2})}{(-\jmath  \frac{2\pi}{\lambda})^l}\notag \\
  &-\frac{e^{-\jmath  \frac{2\pi}{\lambda} f_i(k'd-\frac{3d}{2})}}{-\jmath  \frac{2\pi}{\lambda} f_i^{\prime}(k'd-\frac{3d}{2})} \sum_{l=0}^L \frac{g_i^{(l)}(k'd-\frac{3d}{2})}{(-\jmath  \frac{2\pi}{\lambda})^l}\notag \\
  &+\frac{1}{(-\jmath  \frac{2\pi}{\lambda})^{L+1}} \int_{k'd-\frac{3d}{2}}^{k'd-\frac{d}{2}} e^{-\jmath \frac{2\pi}{\lambda} f_i(t)} g_i^{(L+1)}(t) d t,
\end{align}
with $g_i^{(0)}(t)\triangleq g_i(t)$ and
\begin{align}
  g_i^{(l+1)}(t) = -\frac{d}{d t} \frac{g_i^{(l)}(t)}{f_i^\prime (t)}, \quad l = 0,1,\dots
\end{align}
Here, we can take $L = 0$ and approximate $f(k',i)$ as follows:
\begin{subequations}\label{f_mi}
\begin{align}
  & f(k',i) \notag \\
   &\approx \frac{e^{-\jmath  \frac{2\pi}{\lambda} f_i(k'd-\frac{d}{2})}g_i(k'd-\frac{d}{2})}{-\jmath  \frac{2\pi}{\lambda} f_i^{\prime}(k'd-\frac{d}{2})} \!-\! \frac{e^{-\jmath  \frac{2\pi}{\lambda} f_i(k'd-\frac{3d}{2})}g_i(k'd-\frac{3d}{2})}{-\jmath  \frac{2\pi}{\lambda} f_i^{\prime}(k'd-\frac{3d}{2})}\\
  & = \frac{\jmath \lambda}{2\pi d}(\frac{e^{-\jmath \frac{2\pi}{\lambda}\sqrt{(\Delta_{k',i}+\frac{1}{2})^2 d^2+D^2} }}{\Delta_{k',i}+\frac{1}{2}} \!-\! \frac{e^{-\jmath \frac{2\pi}{\lambda}\sqrt{(\Delta_{k',i}-\frac{1}{2})^2 d^2+D^2} }}{\Delta_{k',i}-\frac{1}{2}} ),
\end{align}
\end{subequations}
where $\Delta_{k',i}\triangleq k'-i$.  Since $f(k',i),\forall k' \neq i$ may be smaller than $f(k',k'),\forall k' \in \cK$,
the terms $f(k',i)f^*(k,i),\forall i \neq k', i \neq k$  can be neglected and the interference terms $\bar{I}_{k',k},\forall k \neq k'$ are approximated as follows:
\begin{align}
  \bar{I}_{k',k}\approx \frac{1}{d^4} \left|f^*(k',k')f(k,k')+f(k',k)f^*(k,k)\right|^2.
\end{align}
By substituting \eqref{f_mm} and \eqref{f_mi}, we obtain the approximated result in \eqref{I_mk}.
\end{proof} 
 
\begin{remark}\label{remark1}
  Although the analysis of inter-user interference is conducted under the worst-case scenario, it is also applicable to more general situations, i.e., $y_m \neq y_i,\forall m,i \in \cK$. This is because the application of the stationary phase method does not put any restrictions on the $y$-coordinates of the users.  Thus, in general cases, the $y$-coordinate can be treated as a constant like $D$, and the analysis process remains the same.
\end{remark}
 
Based on the result in \lemmref{lemma_1}, the system SE under the distributed deployment can be approximated as follows:
\begin{align}\label{SP_approximation}
  C_{\rm d}  \approx \sum_{k=1}^{N} \log_2\left(\!\!1+ \frac{\frac{P_t}{N}\sum_{i=1}^{N} \frac{1}{d_{k,i}^2}}{\sum_{k' \neq k}^{N} \frac{P_t \lambda^3 }{N \pi^2 d^6 D} \frac{ T^2(k,k')}{\sum_{i=1}^{N} \frac{1}{d_{k',i}^2}} + \frac{\sigma^2}{\eta}}\right).
\end{align}
\begin{remark}
  It can be observed from \eqref{SP_approximation} that the main factor that affects the average inter-user interference is the waveguide spacing $d$, since $1/d^6$ is the leading order term. Thus, by letting $d \gg \lambda$, it is possible to make the inter-user interference negligible.
\end{remark}

In addition, since $0 \leq |T(k,k')| \leq 4$, we can obtain the approximate upper and lower bound of $C_{\rm d}$, denoted by $C_{\rm d}^u$ and $C_{\rm d}^l$, which are respectively given by
\begin{subequations}
\begin{align}
  C_{\rm d}^{u} &  =  \sum_{k=1}^{N} \log_2\left(1+\frac{P_t A_k \eta}{N \sigma^2}\right)\label{upper_bound_Cd},\\
  C_{\rm d}^{l}  &= \sum_{k=1}^{N} \log_2\left(1+\frac{\frac{P_t}{N}A_k}{ \sum_{k' \neq k}^{N} \frac{\alpha P_t}{N A_{k'}}+ \frac{\sigma^2}{\eta}}\right)\label{lower_bound_Cd},
\end{align}
\end{subequations}
where $\alpha = \frac{16 \lambda^3}{\pi^2 d^6 D}$, while $A_k = \sum_{i=1}^{N} 1/d_{k,i}^2,\forall k \in \cK$.

\begin{remark}
Note that in practical scenarios, $d$ is much larger than $\lambda$, especially when the signal frequency is high. Thus, it is reasonable to neglect the inter-user interference and approximate $C_{\rm d} $ by $ C_{\rm d}^u$. This implies that that the simple MRT beamformer is sufficient to achieve the maximal system SE. 
This finding will be corroborated by our simulation results presented in the subsequent sections.
\end{remark}

\section{Distributed vs. Centralized Deployment: Multiplexing or Beamforming Gain?}\label{sec: comparison_SE_special_cases}
As pointed before, the distributed deployment would bring high multiplexing gain and the average inter-user interference can be negligible when the waveguide spacing is large enough, while the centralized deployment avails of the high beamforming gain in interference-free transmission. However, it is unknown which deployment strategy is superior.
In this section,  we compare the two deployment strategies of PA systems by evaluating their respective advantages in enhancing the system SE in the high- and low-SNR regimes, respectively.

\subsection{High-SNR Regime}\label{subsec:high multiplexing gain}
As mentioned before, we will use $C_{\rm d}^u$ to approximate $C_{\rm d}$.  Thus, the system SE under the distributed deployment is approximated by
\begin{align}
     C_{\rm d}  \approx \sum_{k=1}^{N} \log_2\left(1+\frac{P_t A_k \eta}{N \sigma^2}\right).
\end{align}
In the high-SNR regime, i.e., when $P_t/\sigma^2 \rightarrow \infty$,  the system SE with the centralized deployment becomes
\begin{subequations}
\begin{align}
C_{\rm c} &\approx \log_2\left(\frac{P_t N \eta}{\sigma^2 d_{k,k}^2}\right) = \log_2\left(\frac{P_t}{\sigma^2}\right)+\log_2\left(\frac{N \eta}{d_{k,k}^2}\right),\\
& \approx \log_2\left(\frac{P_t}{\sigma^2}\right),\label{Cc_app}
\end{align}
\end{subequations}
while the system SE with the distributed deployment is further approximated by
\begin{subequations}
\begin{align}
C_{\rm d} &\approx \sum_{k=1}^{N} \log_2\left(\frac{P_t A_k  \eta }{\sigma^2 N }\right), \\
&= N\log_2\left(\frac{P_t}{\sigma^2}\right) + \sum_{k=1}^{N}\log_2\left( \frac{\eta A_k}{N}\right),\\
& \approx  N\log_2\left(\frac{P_t}{\sigma^2}\right) \label{Cd_app}.
\end{align}
\end{subequations}
The approximations in \eqref{Cc_app} and \eqref{Cd_app} are both because $\log_2(P_t/\sigma^2)$ is the dominant term.
It can be observed that in the high-SNR regime, $C_{\rm d}$ will outperform $C_{\rm c}$ significantly, which shows that the multiplexing gain is more important than the beamforming gain when the SNR is large enough.  Although this is a well-known result in the MIMO community \cite{4200712}, to the best of our knowledge, we are the first to prove it analytically in the PA systems space. Thus, it is suggested to adopt the distributed deployment in order to maximize the system SE in the high-SNR regime.

\subsection{Low-SNR Regime}\label{subsec:high beamforming gain}
In the low-SNR regimes, i.e., when $P_t/\sigma^2 \rightarrow 0$, the system SE with the centralized deployment and the distributed deployment is respectively given by
\begin{subequations}
\begin{align}
  C_{\rm c} &\approx \frac{P_t N \eta }{d_{k,k}^2\sigma^2 \ln2 },\\
  C_{\rm d} &\approx \frac{P_t \eta}{\sigma^2\ln2} \sum_{k=1}^{N} \frac{A_k }{N}.
\end{align}
\end{subequations}
Due to $d_{k,k} \leq d_{k,i},\forall i\neq k$,  we have $A_k/N  = \frac{1}{N}\sum_{i=1}^{N} 1/d_{k,i}^2 \leq 1/d_{k,k}^2$. Meanwhile, $\forall k \in \cK$,  the difference between $1/d_{k,k}^2,\forall k \in \cK,$ can be ignored as no constraints have been imposed on the locations of users around their respective serving waveguides.
 Thus, we have the following inequality:
\begin{align}
  C_{\rm d} \leq \frac{P_t N \eta }{d_{m,m}^2\sigma^2\ln2 } = C_{\rm c}.
\end{align}
It can be found that the centralized deployment is favorable in low-SNR condition to improve the system SE.

The above analysis compares the two deployment strategies in both high- and low-SNR regimes, focusing solely on maximizing the system SE.  To summarize, the centralized deployment is recommended for maximizing the system SE at low SNRs, whereas the distributed deployment performs better in the high-SNR regime.
However,  it is worth noting that the distributed deployment strategy requires $N$ radio-frequency (RF) chains to simultaneously serve $N$ users, which may result in increased hardware complexity and higher energy consumption. This potential drawback is an important factor to consider in practical system design. We will further discuss and evaluate its impact through simulation results in the subsequent section.

\section{Numerical Results}\label{sec: numerical results}
In this section, numerical results are provided to validate the obtained theoretical analysis. Moreover, we compare the system SEs achieved by the centralized deployment and the distributed deployment, and investigate the impact of  waveguide spacing, beamforming strategy, PAs locations, the number of PAs and waveguides.
\subsection{Simulation Setup}
Without additional specifications, we set $N = 5$. All waveguides have the same height of $D = 5$ m, while the length of each waveguide is $L = 10$ m.   The spacing between adjacent waveguides is $d = 2$ m.  The carrier frequency is $f_c = 28$ GHz. The effective refractive index of a dielectric waveguide is $n_{\rm eff} = 1.4$. The power of noise at the receiver is $\sigma^2  = -90$ dBm.    
For simplicity, we set $y_{k'} = y_k, \forall k',k \in \cK$, and the setup of $x_k,k \in \cK$ is given in the system model in \secref{sec:system model}.
To reduce the randomness caused by the distribution of user locations, we generated $100$ times user locations randomly and took the average of the 100 results.
The specific comparison schemes will be detailed in each subsection if necessary.

\subsection{Impact of Approximation}
\begin{figure}[t]
  	\includegraphics[width=3.5in]{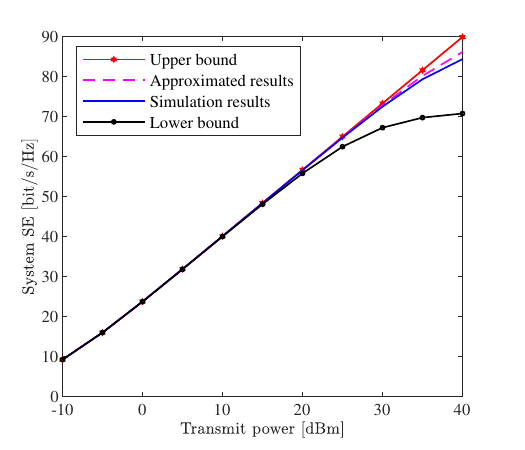}
	\centering
     \captionsetup{font={small}}
	\caption{Comparison between the approximation and the simulation results.}
    \label{comparison}
\end{figure}

In this subsection, we verify the tightness of the approximated results. The ``Simulation results'',``Approximated results'',``Upper bound'', and ``Lower bound'' in \figref{comparison} refer to $C_{\rm d}$ in \eqref{SE_distributed_untractable}, $C_{\rm d}$ in  \eqref{SP_approximation}, $C_{\rm d}^{u}$ in \eqref{upper_bound_Cd} and $C_{\rm d}^{l}$ in \eqref{lower_bound_Cd}, respectively. 
Note that the locations of users are randomly distributed, i.e., $x_k \in [(kd-3d/2,kd-d/2)],\forall k \in \cK$, so that the strength of the inter-user interference considered here is an average value.
To reveal the gap between the approximated results and the simulation results, we consider a relatively small waveguide spacing, i.e., $d = 1$ m, under the distributed deployment.
It can be observed from \figref{comparison} that with the increase of transmit power, the gap between the approximated results and the simulation results is increasing. This is because inter-user interference becomes stronger as the transmit power increases.
However, the gap between the approximated results and the simulation results is overall small even with a large transmit power, which confirms the accuracy of the approximation when evaluating the impact of the inter-user interference.
In addition, it can be observed that the upper bound of $C_{\rm d}$ will continuously increase with the transmit power, since it will not be affected by inter-user interference, while the lower bound of $C_{\rm d}$ will tend to saturate with the increase of transmit power. This is because as the transmit power increases, the system SE is ultimately limited by inter-user interference, which is inherently unavoidable with the MRT beamformer, especially when the spacing $d$ is small.

\subsection{Impact of Deployment Strategies}
\begin{figure}[htbp]
	 \centering
     \captionsetup{font={small}}
     \includegraphics[width=3.5in]{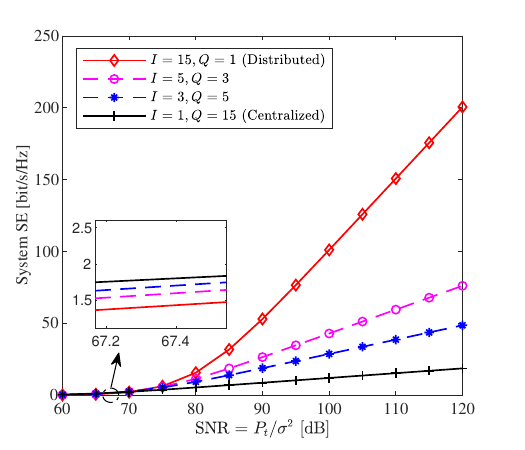}
	 \caption{System SE versus the SNR with varying $I$ and $Q$.}
     \label{comparison_SE}
\end{figure}

\begin{figure}[htbp]
	 \centering
     \captionsetup{font={small}}
     \includegraphics[width=3.5in]{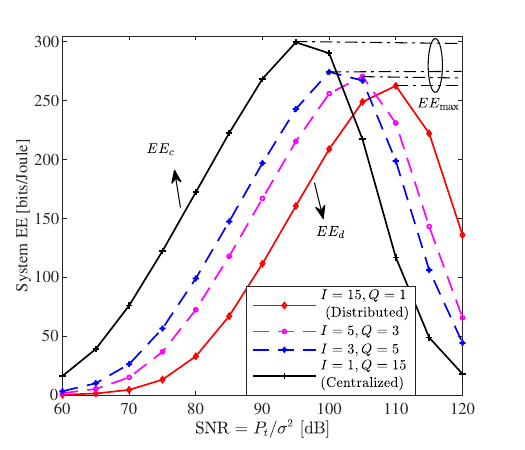}
     \caption{Corresponding system EE versus the SNR with varying $I$ and $Q$.}
     \label{comparison_EE}
\end{figure}
In this subsection, we set $N = 15$ and investigate the impact of different deployment strategies in \figref{comparison_SE} and \figref{comparison_EE}  
to illuminate the trade-off between the multiplexing gain and the beamforming gain. In the general case, $0 \leq I \leq N$ users are served. The number of PAs deployed on each waveguide is $Q$, which ranges between 1 and $N$,  while the total number of PAs in the considered communication system remains constant, i.e., $N = IQ$.  
The resulted system SE in the general case is directly presented as follows:
\begin{align}\label{SE_general_case}
 C  \approx \sum_{k=1}^{I} \log_2\left(\!\!1+ \frac{\frac{P_t}{I}\sum_{i=1}^{I} \frac{1}{d_{k,i}^2}}{\sum_{k' \neq k}^{I} \frac{P_t \lambda^3 }{I \pi^2 d^6 D} \frac{ T^2(k,k')}{\sum_{i=1}^{I} \frac{1}{d_{k',i}^2}} + \frac{\sigma^2}{\eta Q}}\right).
 \end{align}

First, we investigate the impact of the deployment strategies on the system SE in \figref{comparison_SE}.
It can be observed that in the high-SNR regime, the multiplexing gain brings a significant improvement in the system SE, while in the low-SNR regime, the improvement in the system SE brought by the beamforming gain is not obvious. This is because, under the existing waveguide spacing, e.g., $d \geq 2$m, interference between users is negligible, which enables the system SE to achieve its maximum even when using the MRT beamformer.  Therefore, from the perspective of maximizing the system SE, the distributed deployment  is advantageous across most SNR values.

However, it is also worth noting that turning on multiple waveguides means connecting multiple RF chains, which will result in significant power consumptions. Based on the SE analysis in \secref{sec: analysis of system SE},  
let ${\rm EE}_n$ denote the corresponding system EE, where $n$ is the number of the served users in one time slot, varying between 1 and $N$. Note that ${\rm EE}_n$
represents the corresponding EE when the SE is maximized, i.e., the EE achieved when the transmit power is fully utilized.
Moreover, we will mainly consider the energy consumption of RF chains and the transmit power, since the energy consumed by PAs can be considered negligible. Therefore, the power consumption model is
\begin{align}
  P_{\rm con} = nP_{\rm RF} + P_t,
\end{align}
where $n=1$ with the centralized deployment and $n=N$ with the distributed deployment. The power consumption per RF chain is $P_{\rm RF} = 31.6$ mW \cite{castellanos2025embracing}. Therefore, the EE of the PA-based system is
\begin{align}
  {\rm EE}_n = \frac{C_n}{P_{\rm con}},
\end{align}
where $C_n$ is the system SE depending on the deployment strategies. Specifically, $C_N = C_{\rm d}$ with the distributed deployment and $C_1 = C_{\rm c}$ with the centralized deployment.

It can be observed from \figref{comparison_EE} that ${\rm EE}_N$ is significantly smaller than ${\rm EE}_1$ in the low-SNR regime. 
Furthermore, we have also noticed that the maximum system EE decreases as $I$ increases, while more transmit power is needed to  reach the maximal system EE. 
Thus, from the perspective of energy consumption, the centralized deployment is always advantageous in the low-SNR regime.
Although we focus on comparing the advantages of the two deployment strategies in maximizing the SE, the EE is also an essential factor that must be considered in practical applications.
The above analysis reveals that a SE-maximizing deployment favors the distributed deployment, while an EE-maximizing deployment leans toward the centralized deployment. Therefore, in practical applications, careful selection of the $I$ and $Q$ components is essential to strike a good trade-off between the SE and EE.

\subsection{Impact of  Waveguide Spacing}
\begin{figure}[t]
	\centering
     \captionsetup{font={small}}
     \subfigure[$P_t = 0$ dBm]{\includegraphics[width=3.5in]{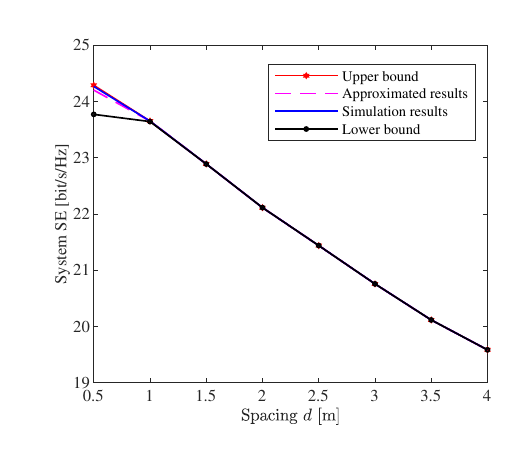}}
     \subfigure[$P_t = 40$ dBm]{\includegraphics[width=3.5in]{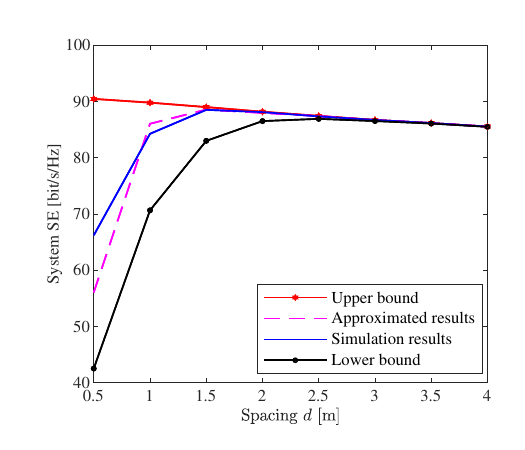}}
	 \caption{System SE versus the spacing of adjacent waveguides.}
    \label{impact_d}
\end{figure}

In this subsection, we investigate the impact of the waveguide spacing in \figref{impact_d}, i.e., the impact of $d$, with the transmit power being $P_t = 0$ dBm and $P_t = 40$ dBm, respectively. Here, we only consider the distributed deployment since there is no interference with the centralized deployment.
It can be observed from \figref{impact_d} that with the increase of transmit power, the system SE will eventually reach its upper bound in both cases. This is because with the increase of $d$, the inter-user interference will become negligible. Moreover, the results in \figref{impact_d} show that it is reasonable to approximate the system SE with the upper bound when $d$ is large, e.g., $d \geq 2$ m in practical applications. 
Meanwhile, the system SE nearly reaches the upper bound when $P_t = 0$ dBm, while $P_t = 40$ dBm, it gradually approaches the upper bound as $d$ increases.
This also implies that when operating at high SNRs, we may need a larger $d$.
In addition, while the lower bound of the system SE is growing larger with increasing $d$, the upper bound is decreasing with increasing $d$. The former results from the reduced inter-user interference, while the latter stems from the increased path loss due to larger waveguide spacing. Thus, increasing $d$ is a double-edged choice: while larger $d$ helps render the inter-user interference negligible, it may also entail increasing propagation loss induced by larger propagation distance. This indicates that there exists an optimal waveguide spacing in practical applications, which could be a topic of follow-up research.

\subsection{Impact of Beamforming Strategy}
\begin{figure}[t]
	\centering
    \captionsetup{font={small}}
    \includegraphics[width=3.5in]{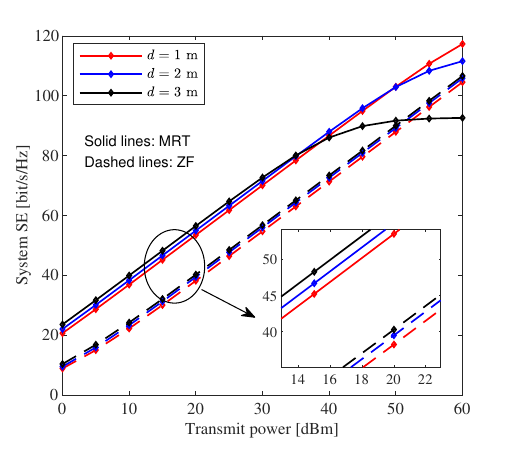}
	\caption{System SE with different beamformers.}
    \label{impact_beamformer}
\end{figure}
In this subsection, we investigate the impact of beamforming strategy under the distributed deployment in \figref{impact_beamformer}, since the inter-user interference is the most severe in this case. 
Here, the ZF beamformer is adopted as a baseline, which is $\bW^T = \sqrt{\alpha}\bH^H(\bH\bH^H)^{-1} $, with $\alpha = N/\trace{(\bH\bH^H)^{-1}}$, $\bH  = [\bh(\br_1,\bs),\bh(\br_2,\bs),\dots, \bh(\br_N,\bs)]$  and $\bW = [\bw_1,\bw_2,\dots,\bw_N]$.
Thus, the system SE is given by
\begin{align}
  C^{\rm ZF} = \sum_{k=1}^{N} \log_2\left(1+ \frac{\alpha P_k}{\sigma^2}\right) = N \log_2\left(1+ \frac{\alpha P_t}{N\sigma^2}\right).
\end{align}

It can be observed from \figref{impact_beamformer} that the system SE achieved by the MRT beamformer exceeds that of the ZF beamformer over most of the transmit power range, especially when $d$ is large. 
This is because larger $d$ makes the inter-user interference negligible. 
Thus, by using the MRT beamformer, the strength of the received signal is able to be maximized, while the ZF beamformer fails to maximize the received signal strength, as it aims for the complete elimination of inter-user interference.  However, in the high-SNR regime with small waveguide spacing, the SE achieved by MRT underperforms that with ZF. This occurs because inter-user interference becomes non-negligible in this scenario. Under these conditions, ZF's inherent capability to completely eliminate inter-user interference provides a distinct advantage.
Moreover, it can be observed that during the linear growth stage of the system SE achieved by the MRT beamformer with respect to the transmit power, an increase in $d$ leads to a slight degradation in the system SE. This occurs because a larger $d$ corresponds to larger signal propagation distances, resulting in larger path loss, which aligns with the phenomenon observed in \figref{impact_d}. The same behavior is observed for the SE under the ZF beamformer.

\subsection{Impact of PAs Locations}
\begin{figure}[htbp]
	\centering
  	\subfigure[Different location strategies of PAs with the centralized deployment.]{\includegraphics[width=3.5in]{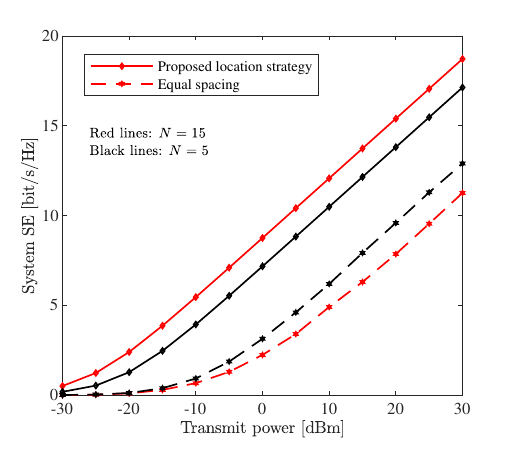}\label{locations_PAs_centralized}}
    \subfigure[Different location strategies of PAs with the distributed deployment.]{\includegraphics[width=3.5in]{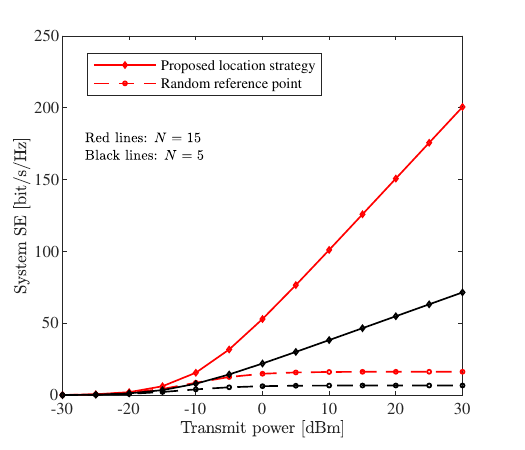}\label{locations_PAs_distributed}}
     \captionsetup{font={small}}
	\caption{System SE with different location strategies of PAs.}
    \label{impact_positions}
\end{figure}
In this subsection, we compare the different location strategies of the PAs on the waveguides under the centralized deployment and the distributed deployment, respectively. The baselines are set as follows:
\begin{itemize}
\item \textbf{Equal spacing:} To investigate the impact of the locations of PAs on the same waveguide under the centralized deployment, let the PAs be equally spaced on the waveguide, i.e.,  the spacing between adjacent PAs is $d_s = L/(N-1)$.
\item \textbf{Random reference point:} To investigate the impact of the locations of PAs on the waveguides under the distributed deployment, let the spacing between the adjacent PAs on the same waveguide remain $\lambda_e$ to ensure that signals radiated by different PAs on the same waveguide are in phase, while the reference point is randomly selected in this baseline scheme. Thus, the common phase factor is random here, which will influence the strength of the inter-user interference under the distributed deployment.
\end{itemize}
The ``Proposed location strategy'' in \figref{impact_positions} corresponds to the location strategies mentioned in \subsecref{SE_Analysis_Centralized} and \subsecref{SE_Analysis_Distributed}, where the spacing between adjacent PAs on the same waveguide is $\lambda_e$, with the reference point being the location on the waveguide closest to the user. 
First, it can be observed from \figref{locations_PAs_centralized} that the proposed location strategy outperforms the equal spacing baseline, since the former ensures the signals sent from waveguide $k$ to user $k,\forall k \in \cK$  are maximized. 
In addition, it can be noticed that while the number of PAs increases from $N=5$ to $N=15$, the system SE achieved by the equal spacing baseline decreases, which is different from that in the proposed location strategy. Note that in the equal spacing baseline, the system SE under the centralized distributed deployment is given as follows: 
\begin{align}
  C_{\rm e} =\frac{1}{N} \sum_{k=1}^{N} \log_2\left(1+ \frac{P_t \eta}{\sigma^2} L_e\right) \overset{\text{(a)}}{\approx}\log_2\left(1+ \frac{P_t \eta}{\sigma^2} L_e\right),
\end{align}
where $L_{\rm e} \triangleq \frac{1}{N}\big|\sum_{q=1}^{N} \frac{1}{d_{k,k,q}}\big|^2 \neq  \frac{N}{d_{k,k}^2}$, since the spacing between adjacent PAs is not negligible. The approximation in (a) is because the distribution of $\{d_{k,k,q}\}_{q=1}^N$ can be considered identical for any $k \in \cK$.
The value of $L_{\rm e}$ is dependent on the users' locations as well as the PAs' locations on the waveguide. Due to larger propagation distance-induced path loss, $C_{\rm e}$ decreases when $N$ changes from 5 to 15.
This also demonstrates that the deployment of PAs on the waveguide is crucial for the system SE under the centralized deployment strategy.

On the other hand, it can be observed from \figref{locations_PAs_distributed} that the proposed location strategy also achieves higher SE than the random reference point baseline. This is because the former minimizes the large-scale path loss of the received signal by deploying the reference PA closest to the user.

\subsection{Sensitivity Analysis of $I$ and $Q$}
\begin{figure}[htbp]
	\centering
  	\subfigure[$P_t = 0$ dBm]{\includegraphics[width=2.5in]{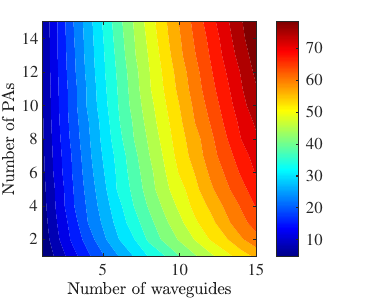}\label{sensitivity_0dBm}}
    \subfigure[$P_t = 40$ dBm]{\includegraphics[width=2.5in]{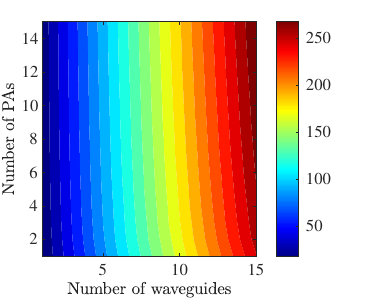}\label{sensitivity_40dBm}}
     \captionsetup{font={small}}
	\caption{Sensitivity of $I$ and $Q$ and their impact on the system SE.}
    \label{sensitivity_analysis}
\end{figure}
In this section, we analyze the sensitivity of the system SE to the parameters $I$ and $Q$ under different transmit power levels. Regions with the same color in the figure represent identical system SE. It can be observed that there exist multiple combinations of $I$ and $Q$ that achieve the same SE. Moreover, the system SE is more sensitive to variations in $I$ compared to $Q$. Additionally, under high transmit power, i.e., $P_t = 40$ dBm, the sensitivity of system SE to $Q$ is reduced. This is reflected in the fact that placing one or more PAs on the waveguide yields nearly identical system SE at high transmit power. This observation provides a practical insight for PA deployment: under high transmit power conditions, it is sufficient to place only one PA on the waveguide to keep the overall cost to affordable levels. Furthermore, when optimizing the locations of PAs in the PA-based system design, assigning only one PA to each waveguide will significantly reduce the optimization complexity.

\section{Conclusions}\label{sec: conclusions}
In this paper, we investigated the centralized and distributed deployment strategies in PA-based multi-user communication systems, respectively, revealing their advantages in enhancing the system SE.
First, we established the system models along with the SE models under the two deployment strategies. For the centralized PA deployment, we determined the locations of PAs on the waveguide first and then obtained the system SE directly. For the distributed PA deployment, we determined the locations of PAs on each waveguide and obtained the system SE by utilizing MRT beamformer. After that, we leveraged the stationary phase method and approximated the system SE under the distributed deployment with an analytically tractable form. By analyzing the approximated result, it was found that by enlarging the waveguide spacing, the average inter-user interference could be assumed as negligible.
Based on the derived system SEs under the two deployments, we compared them in the high- and low-SNR regimes, respectively. This analysis revealed that in the high-SNR regime, the distributed deployment strategy was more effective in maximizing the system SE, whereas in the low-SNR regime, the centralized deployment was more advantageous.
Finally, we validated all the analytical results through simulations. It was demonstrated that \romannumeral1) by applying the stationary phase method, the system SE can be accurately approximated;
\romannumeral2) although a large waveguide spacing  renders the inter-user interference negligible, making MRT sufficient to maximize the system SE under the distributed deployment,  it also led to higher propagation distance-induced path losses; and \romannumeral3) while the distributed deployment was favorable  for maximizing the system SE, the centralized deployment was preferable for maximizing the system EE. Therefore, to achieve both high SE and EE, a hybrid architecture might have to be customized.

\bibliographystyle{IEEEtran.bst}
\bibliography{Refabrv_20180802,ref_pinch.bib}

\end{document}